\documentclass[runningheads,fleqn]{svmult}

\usepackage{makeidx}   
\usepackage{graphicx}  
\usepackage{subeqnar}  
\usepackage{multicol}  
\usepackage{physmult}  
\makeindex             

\begin{document}
\title*{Self-Consistency\\ of Thermal Jump Trajectories}
\toctitle{Self-Consistency of Thermal Jump Trajectories}
%
%
%
\author{Y.-T. Chough \and H. J. Carmichael}

%
%
%

\maketitle              

\abstract{It is problematic to interpret the quantum jumps of an atom
interacting with thermal light in terms of counts at detectors
monitoring the atom's inputs and outputs. As an alternative, we develop
an interpretation based on a self-consistency argument. We include one
mode of the thermal field in the system Hamiltonian and describe its
interaction with the atom by an entangled quantum state while assuming
that the other modes induce quantum jumps in the usual fashion. In the
weak-coupling limit, the photon number expectation of the selected mode
is also seen to execute quantum jumps, although more generally, for
stronger coupling, Rabi oscillations are observed; the equilibrium photon
number distribution is a Bose-Einstein distribution. Each mode may be
viewed in isolation in a similar fashion, and summing over their weak-coupling
jump rates returns the net jump rates for the atom assumed at the outset.}

\section{Introduction}
The notion of a quantum jump entered physics with Bohr's model of the
atom and was elaborated in a semi-quantitative formulation in Einstein
$A$ and $B$ theory \cite{einstein16}. It was, from the beginning, an idea
at variance with the usual commitment to a continuous time evolution
and with the continuous constitution of light as an electromagnetic wave.
The appearance of the Schr\"odinger equation relieved the situation
somewhat, but ultimately, through the use of perturbation theory to
make testable predictions about quantum scattering processes, the quantum
jump remains with us, though certainly in a more sophisticated and
mathematically refined form.

The Monte-Carlo wavefunction and quantum trajectory methods developed in
quantum optics \cite{dalibard92,dum92,carmichael93} use quantum jumps
as an explicit component of a stochastic time evolution in a manner very
reminiscent of Einstein $A$ and $B$ theory. In fact, the only novelty is
to combine Einstein's rules for quantum jumps with a coherent evolution
between jumps that admits a dynamic involving superpositions of stationary
states. In this, these methods achieve something remarkably similar to the
proposal of Bohr, Kramers, and Slater (BKS) \cite{bohr24,slater24} for
uniting discontinuous jumps among the stationary states of a material
system with a continuous evolution between jumps, during which time
material oscillators, possessing coherent amplitudes, are brought into play.

The realistic interpretation sought by BKS may not, however, be entertained.
In most circumstances, an interpretation of the jumps employed in quantum
trajectory theory is based upon a record of time-resolved photon
counts which might be realized in practice by terminating every output
channel in a photodetector \cite{carmichael93,carmichael99}. This scenario
is plausible because optical frequencies are sufficiently high that
scattered photons can be detected against an essentially vacuum-state
background. The measurement-based interpretation is problematic, though,
for an atom exchanging photons with a thermal environment. In this situation,
incoherent photons are both emitted {\it and\/} absorbed; moreover, it
is impossible to distinguish a scattered photon from some other photon in
the environment. Of course, schemes such as electron shelving exist that are
able to monitor thermal quantum jumps \cite{nagourney86,sauter86,bergquist86}.
They, however, make intrusive measurements by utilizing strong couplings to
other inputs and outputs, and are not a suitable foundation for the
interpretation of quantum trajectory equations. The relationship, in fact,
is exactly the reverse; electron shelving is one of the measurement schemes
that quantum trajectories would propose to explain.

In this paper we follow a different direction to give substance, beyond a
mere assertion, to the interpretation that an atom exchanging photons
with a thermal environment does, in some well-defined sense, execute
jumps. We define the sense through a self-consistency argument. Contrary
to the notion of quantum jumps, quantum mechanics continuously entangles
two interacting systems through the Schr\"odinger evolution. We show that
a single mode, selected from the many modes of a thermal environment,
and allowed to evolve in interacting with an atom to produce such an
entanglement, is, in fact, seen to undergo jumps in its photon number
expectation if it is assumed that all other modes of the environment,
treated collectively as a reservoir, induce jumps between the atomic
states according to the rules of Einstein theory. The jump evolution for
the selected mode emerges from an otherwise continuous evolution in the
weak-coupling limit. Thus, assuming jumps induced by the reservoir as a
whole leads, self-consistently, to the appearance of jumps in an
individual mode of the reservoir when that mode is allowed to entangle
with the atom {\it via} the standard interaction Hamiltonian and
Schr\"odinger evolution. The jumps do bring the individual mode to a 
Bose-Einstein distribution over photon number, and the jump rates, derived
for every mode viewed individually in this way, sum to net rates which
agree with Einstein $A$ and $B$ theory.
 
The underlying theme of this paper is the conflict between a continuous
and a discontinuous quantum evolution and the self-consistency of the two
in the perturbative weak-coupling limit. We therefore briefly review, in
Sect.~\ref{sec:two}, the quantum jump model of Einstein, the BKS proposal
to include a continuous evolution, and the quantum trajectory
realization of the latter. The argument for the self-consistency of
thermal jump trajectories is elaborated in Sect.~\ref{sec:three}.
\section{Einstein $A$ and $B$ Theory, BKS, and Quantum Trajectories}
\label{sec:two}
\subsection{Thermal Quantum Jumps}
\begin{figure}
\sidecaption
\includegraphics[width=.3\textwidth]{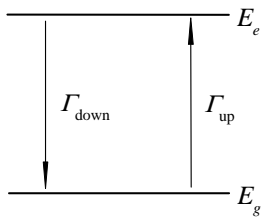}
\caption[]{Thermal quantum jumps in Einstein $A$ and $B$ theory.
The jump rates are defined in (\ref{eqn:onea}) and (\ref{eqn:oneb}).}
\label{fig:one}
\end{figure}

We consider the single two-state atom illustrated in Fig.\ \ref{fig:one},
in thermal equilibrium with Planck radiation at temperature $T$. In
Einstein $A$ and $B$ theory photons are exchanged between the atom and the
radiation field as the atom jumps randomly between its two stationary
states. The jump rates follow a prescription taking into account spontaneous
emission, stimulated emission, and absorption, with \cite{einstein16}
\begin{subeqnarray}
\Gamma_{\rm down}&=&A+B\sigma(\omega_0)\;,
\label{eqn:onea}\\
\Gamma_{\rm up}&=&B\sigma(\omega_0)\;,
\label{eqn:oneb}
\end{subeqnarray}
where 
\begin{equation}
\sigma(\omega_0)=\bar n(\omega_0)\hbar\omega_0[\rho(\omega_0)/V]
\end{equation}
is the energy density of the radiation field at the resonance frequency
$\omega_0$ of the atom, with average photon number per mode
\begin{equation}
\bar n(\omega_0)=[e^{\hbar\omega_0/k_BT}-1]^{-1}
\end{equation}
and mode density (in volume $V$)
\begin{equation}
\rho(\omega_0)=\frac{\omega_0^2V}{\pi^2c^3}\;,
\end{equation}
and the Einstein $A$ and $B$ coefficients must satisfy
\begin{equation}
\frac BA=\frac{\pi^2c^3}{\hbar\omega_0^3}
\label{eqn:five}
\end{equation}
in order for the atom to be brought into thermal equilibrium with the
radiation.

With the help of the relationship (\ref{eqn:five}), we write
(\ref{eqn:onea}) and (\ref{eqn:oneb}) in the modern notation
\begin{subeqnarray}
\Gamma_{\rm down}&=&A[\bar n(\omega_0)+1]\;,
\label{eqn:sixa}\\
\Gamma_{\rm up}&=&A\bar n(\omega_0)\;.
\label{eqn:sixb}
\end{subeqnarray}
Einstein theory does not assign a value to the coefficient $A$. From quantum
mechanics, however, using Fermi's Golden rule we obtain
\begin{equation}
A=2\pi\sum_\lambda\int\! d\Omega\,\rho(\omega_0)|\kappa_{\lambda,\hat{\vec n}}
(\omega_0)|^2\;,
\label{eqn:seven}
\end{equation}
where
\begin{equation}
|\kappa_{\lambda,\hat{\vec n}}(\omega_0)|=\sqrt{\frac{\omega_0}
{2\pi\epsilon_0 V}}|\hat e_{\lambda,\hat{\vec n}}\cdot\vec d_{eg}|
\label{eqn:eight}
\end{equation}
is the dipole coupling strength to a mode of the radiation field with
polarization $\lambda$ and direction of propagation specified by the unit
vector $\hat{\vec n}$ (polarization vector $\hat e_{\lambda,\hat{\vec n}}$);
$\vec d_{eg}$ is the atomic dipole matrix element.

\begin{figure}
\sidecaption
\includegraphics[width=.6\textwidth]{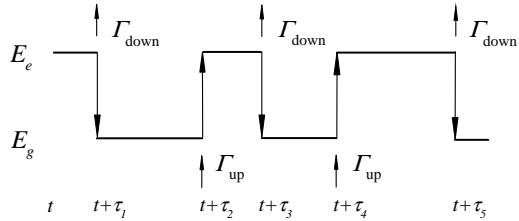}
\caption[]{Sample realization of the Einstein stochastic process.
The corresponding record is defined in (\ref{eqn:record}).}
\label{fig:two}
\end{figure}

Commonly, Einstein theory is discussed at the level of rate equations for
the occupation probabilities of the atomic stationary states. The theory
does, however, define a stochastic process -- one that may be visualized
in terms of quantum jumps whose occurrences unfold randomly in time. With
each realization of the stochastic process we associate a record of jump
types and jump times,
\begin{equation}
{\rm REC}\equiv\left\{\matrix{\ldots&\Gamma_{\rm down}&\Gamma_{\rm up}&
\Gamma_{\rm down}&\Gamma_{\rm up}&\Gamma_{\rm down}&\ldots\cr
\ldots&t+\tau_1&t+\tau_2&t+\tau_3&t+\tau_4&t+\tau_5&\ldots}\right.\;.
\label{eqn:record}
\end{equation}
In Fig.~\ref{fig:two} we illustrate the discontinuous evolution of the
atom in coordination with its absorption and emission of thermal photons.

\subsection{Coherence: The BKS Proposal}
Although Bohr had himself put forward the notion of a quantum jump to explain
the association of stationary-state energy differences with electromagnetic
wave frequencies in his model of the hydrogen atom, he became quite
dissatisfied with the idea in the concrete form it acquired under Einstein's
proposal. It was the light quantum, specifically, that troubled him the most.
Bohr insisted that since so many optical phenomena rely on the continuity
of coherent waves, the wave nature of light simply could not be dismissed.
He recognized, on the other hand, that a discontinuous process was definitely
needed to account for light emission and detection. What he was unavoidably
drawn towards, then, was some sort of merging of the two ideas.

A program to accomplish this was outlined in what has come to known as the
BKS proposal \cite{bohr24,slater24}. The central idea of this proposal is the
proposition that during the residence times in a stationary state, represented
by the horizontal lines in Fig.~\ref{fig:two}, an atom is not inactive in its
interaction with the electromagnetic field; rather, it acts through a coherent
dipole radiator, or ``virtual oscillator'' in the words of BKS, which is all
the time radiating an electromagnetic wave of frequency $\omega_0=(E_e-E_g)/
\hbar$. This wave is either in phase or out of phase with the external radiation
at frequency $\omega_0$ depending on whether the residence is in the stationary
state with energy $E_e$ or $E_g$. Thus, there is an energy transfer under the
laws of classical electrodynamics either to the electromagnetic field from
the dipole, or in the reverse direction, depending on the stationary state
\cite{note1}. Clearly a double counting of the exchanged energy occurs if one
accepts that light quanta are also emitted and absorbed at the times of the
jumps. However, the goal was precisely to eliminate these quanta, although
still permitting the atom to jump. BKS attempted, thus, to retain, 
but keep separate, two incompatible mechanisms for energy exchange -- a
wave (continuous) mechanism for the absorption and emission of radiation and a
particle (discontinuous) mechanism for the change of material state energies.
Their proposal foundered on its obvious violation of energy conservation at
the level of the individual quantum events, a feature that appeared not to be
supported in Compton scattering experiments \cite{bothe25,compton25}.

\subsection{Coherence: Quantum Trajectories}
In retrospect we can see that Bohr had in mind a conception of light that,
although supported by numerous wave phenomena in optics, was largely inappropriate
for the light sources available at the time. Radiators in thermal equilibrium are
not sources of coherent waves. They radiate electromagnetic noise, which, with
filtering, can approximate {\it low intensity\/} light possessing first-order
coherence, but is very far from the classical concept of a coherent wave of
large and adjustable amplitude. Modern lasers, however, emit something close to
the classical ideal. In their case, the high coherence, if it is to be preserved,
disallows the tracking of energy at the level of the individual quanta, so that
the energy conservation argument against the BKS proposal does not apply.

Quantum trajectory theory is designed to deal with problems involving the
interaction of matter with the high coherence light sources available in modern
laboratories. It shows remarkable similarities to the BKS proposal. These are
described elsewhere \cite{carmichael97,carmichael99}, and we do not plan to
discuss them in any depth here. As an introduction to our main topic, however,
it is useful to contrast Einstein $A$ and $B$ theory with the quantum
trajectory description of a coherent field, amplitude ${\cal E}$, resonantly
exciting the two-state atom of Fig.~\ref{fig:one}. The connections with BKS
emerge automatically through this exercise.

The atom is still located in a thermal environment and quantum jumps still
appear as they do in Fig.~\ref{fig:two} -- but with one notable modification.
Due to the induced coherence, it is necessary that the system state be a
superposition of the stationary states $|E_e\rangle$ and $|E_g\rangle$,
which we denote by $|\psi_{\rm REC}(t)\rangle$. As suggested by BKS, there
is a coherent interaction with the electromagnetic field between the
quantum jumps. This we account for by a continuous evolution under the
Schr\"odinger equation \cite{dalibard92,dum92,carmichael93} (for the
unnormalized conditional state)
\begin{equation}
\frac{d|\bar\psi_{\rm REC}\rangle}{dt}=\frac1{i\hbar}\hat H_B
|\bar\psi_{\rm REC}\rangle\;,
\label{eqn:ten}
\end{equation}
with non-Hermitian Hamiltonian
\begin{eqnarray}
\hat H_B&=&{\textstyle{1\over2}\displaystyle}\hbar(\omega_0-i\Gamma_{\rm down}
)|E_e\rangle\langle E_e|-{\textstyle{1\over2}\displaystyle}\hbar(\omega_0
+i\Gamma_{\rm up})|E_g\rangle\langle E_g|\cr
\noalign{\vskip3pt}
&&+i\hbar{\cal E}(e^{i\omega_0t}|E_g\rangle\langle E_e|-e^{-i\omega_0t}
|E_e\rangle\langle E_g|)\;,
\label{eqn:eleven}
\end{eqnarray}
in which the external coherent field is classical, and its interaction with the
atom is treated in the dipole and rotating-wave approximations. The quantum
jumps are governed by the probabilistic rules of Einstein $A$ and $B$ theory,
generalized, in a natural way, to account for the fact that the system at any
time is not definitely in a particular stationary state. There are jumps
\begin{subeqnarray}
|\bar\psi_{\rm REC}\rangle&\raise5pt\hbox{$\matrix{{}_{\Gamma_{\rm down}}\cr
\noalign{\vskip-3pt}\to}$}&(|E_g\rangle\langle E_e|)|\bar\psi_{\rm REC}\rangle\;,\\
\noalign{\vskip-2pt}
|\bar\psi_{\rm REC}\rangle&\raise5pt\hbox{$\matrix{{}_{\Gamma_{\rm up}}\cr
\noalign{\vskip-3pt}\to}$}&(|E_e\rangle\langle E_g|)|\bar\psi_{\rm REC}\rangle\;,
\end{subeqnarray}
with jump rates
\begin{subeqnarray}
R_{\rm down}&=&\Gamma_{\rm down}|\langle E_e|\psi_{\rm REC}\rangle|^2\;,\\
R_{\rm up}&=&\Gamma_{\rm up}|\langle E_g|\psi_{\rm REC}\rangle|^2\;.
\label{eqn:thirteenb}
\end{subeqnarray}

\begin{figure}
\vbox{\vskip-1.5cm}
\includegraphics[width=.96\textwidth]{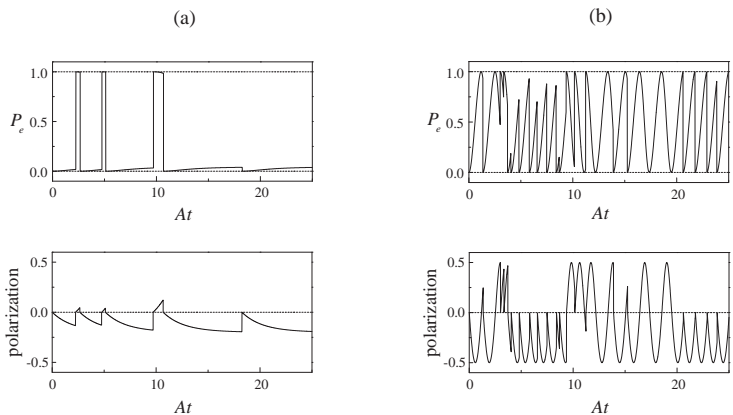}
\caption[]{Sample quantum trajectories with both thermal jumps and
induced coherence: $\bar n(\omega_0)=0.25$, ${\cal E}/A=0.1$ and
$1.5$ [({\bf a}) and ({\bf b})].}
\vbox{}
\label{fig:three}
\end{figure}

Figure \ref{fig:three} illustrates how realizations of this stochastic process
appear. In (a) the coherent excitation is relatively weak and the overall form of
the evolution remains close to that produced by the Bohr-Einstein quantum jumps
(Fig.~\ref{fig:two}). There is, however, in addition to the switching of the
energy, a weak induced coherence carried along by the continuous evolution between
jumps as a nonvanishing polarization amplitude, quite reminiscent of the BKS
virtual oscillator. In Fig.~\ref{fig:three}(b) the coherent excitation is much
stronger. Here, the dominant mechanism for evolution between the stationary
states is a coherent Rabi oscillation; we move to a coherent evolution that
is nonperturbative and a regime which only became accessible with the invention
of the laser. This, specifically, is a form of evolution following from the
Schr\"odinger equation; {\it nonperturbative\/} coherence was not anticipated
by the BKS proposal. We might note in passing that lasing without inversion
and related phenomena acquire their counterintuitive features from such
nonperturbative coherence \cite{carmichael97}.

\section{Self-Consistency of Thermal Quantum Jumps}
\label{sec:three}
At a sufficiently low temperature, when $\bar n(\omega_0)\ll1$, the jump record
that labels the state can reasonably be made by detectors monitoring the scattered
light. Almost all jumps will be down-jumps governed by the spontaneous emission
rate $A$ which may be identified with emitted photons counted as isolated (in space
and time) excitations of the vacuum. For many applications in quantum optics this
is the situation in reality. Nonetheless, the thermal jumps, though they might
be negligible in practice, cannot be set aside from a fundamental point of view.
Considering then those jumps which cannot reasonably be identified with the
``click'' of a detector, is there any other justification for returning to the
language of the old quantum theory, given that, in quantum mechanics, the
Schr\"odinger equation invokes only a continuous evolution? We aim to show that
there is, in so far as the jumps are self-consistent -- consistent with the
Schr\"odinger evolution -- in the weak-coupling limit.  

\subsection{Trajectories for a Single Field Mode}
The Hamiltonian for a two-state atom interacting with the radiation field
of a thermal environment (we now take ${\cal E}=0$) is
\begin{eqnarray}
\hat H&=&{\textstyle{1\over2}\displaystyle}\hbar\omega_0|E_e\rangle\langle E_e|
-{\textstyle{1\over2}\displaystyle}\hbar\omega_0|E_g\rangle\langle E_g|\cr
\noalign{\vskip4pt}
&&+\sum_{\lambda^\prime,\hat{\vec n}^\prime,\omega^\prime}\hbar\omega^\prime
\hat r^\dagger_{\lambda^\prime,\hat{\vec n}^\prime,\omega^\prime}
\hat r_{\lambda^\prime,\hat{\vec n}^\prime,\omega^\prime}\cr
\noalign{\vskip2pt}
&&+\sum_{\lambda^\prime,\hat{\vec n}^\prime,\omega^\prime}
\hbar[\kappa_{\lambda^\prime,\hat{\vec n}^\prime}(\omega^\prime)
|E_e\rangle\langle E_g|\hat r_{\lambda^\prime,\hat{\vec n}^\prime,\omega^\prime}
+{\rm H.c.}]\;,
\label{eqn:fourteen}
\end{eqnarray}
where $\hat r^\dagger_{\lambda,\hat{\vec n},\omega}$ and
$\hat r_{\lambda,\hat{\vec n},\omega}$ are creation and annihilation
operators for the field mode with polarization $\lambda$, propagation
direction $\hat{\vec n}$, and frequency $\omega$, and
$\kappa_{\lambda,\hat{\vec n}}(\omega)$ is the mode coupling coefficient
whose magnitude is defined in (\ref{eqn:eight}). The stochastic process
(\ref{eqn:ten})--(\ref{eqn:thirteenb}) (${\cal E}=0$) is developed, formally,
around the master equation derived from (\ref{eqn:fourteen})
\cite{dalibard92,dum92,carmichael93}. This master equation describes the
quantum state of the atom alone, after tracing over every mode of the
radiation field. Our idea is to raise one mode of the field to the same status
as the atom by including it, along with its interaction with the atom, in the
system Hamiltonian. All other modes are to be treated as a reservoir as
before and their interaction with the atom described by quantum jumps. The
stochastic process is the same as in (\ref{eqn:ten})--(\ref{eqn:thirteenb}),
but with the non-Hermitian Hamiltonian $\hat H_B$ replaced by
\begin{eqnarray}
\hat H_B&=&{\textstyle{1\over2}\displaystyle}\hbar(\omega_0-i\Gamma_{\rm down})
|E_e\rangle\langle E_e|-{\textstyle{1\over2}\displaystyle}\hbar(\omega_0+
i\Gamma_{\rm up})|E_g\rangle\langle E_g|\cr
\noalign{\vskip1pt}
&&+\hbar\omega\hat r^\dagger_{\lambda,\hat{\vec n},\omega}
\hat r_{\lambda,\hat{\vec n},\omega}\cr
\noalign{\vskip3pt}
&&+\hbar[\kappa_{\lambda,\hat{\vec n}}(\omega)|E_e\rangle\langle E_g|
\hat r_{\lambda,\hat{\vec n},\omega}+{\rm H.c.}]\;.
\label{eqn:fifteen}
\end{eqnarray}
Of course removing one mode from the reservoir has no effect on the overall
jump rates for the atom. The change is that we can now follow the evolution
of an explicit Hilbert space vector for the selected mode, one that entangles
this mode with the atom. We ask how does the selected mode evolve in the
Hilbert space; in particular, does it also experience quantum jumps?

\begin{figure}
\vbox{\vskip-2.5cm}
\sidecaption
\includegraphics[width=.55\textwidth]{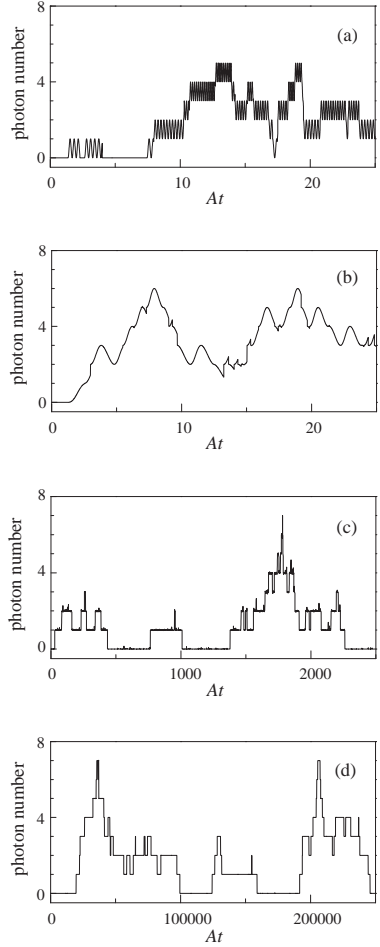}
\caption[]{Sample trajectories for the photon number expectation
of a single field mode included as part of the system. The mode is
resonant: $\Delta\omega/A=0$, $\bar n(\omega_0)=1$, and
$|\kappa_{\lambda,\hat{\vec n}}(\omega_0)|/A=10$, $1.0$, $0.1$,
and $0.01$ [({\bf a}), ({\bf b}), ({\bf c}), and ({\bf d})].\par
\vbox{\vskip0.25cm}}
\label{fig:four}
\end{figure}

\begin{figure}
\vbox{\vskip-2.5cm}
\sidecaption
\includegraphics[width=.55\textwidth]{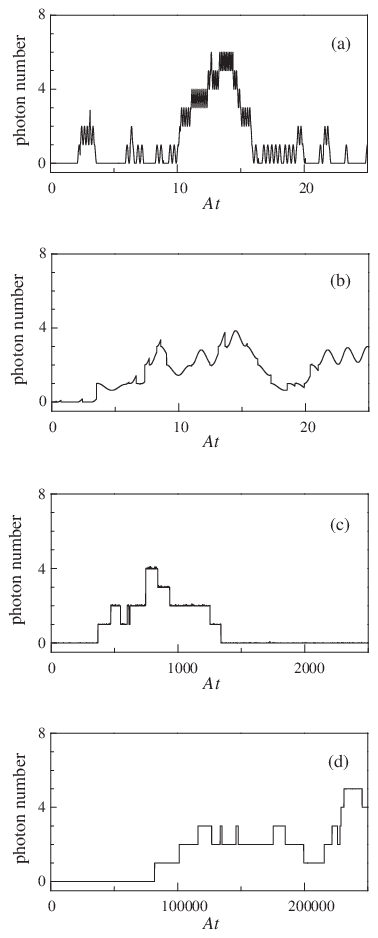}
\caption[]{Sample trajectories for the photon number expectation
of a single field mode included as part of the system. The mode is
nonresonant: $\Delta\omega/A=2.0$, $\bar n(\omega_0)=1$,
and $|\kappa_{\lambda,\hat{\vec n}}(\omega)|/A=10$, $1.0$, $0.1$,
and $0.01$ [({\bf a}), ({\bf b}), ({\bf c}), and ({\bf d})].\par
\vbox{\vskip0.25cm}}
\label{fig:five}
\end{figure}

Figures~\ref{fig:four} and \ref{fig:five} show sample trajectories for the
selected-mode photon number expectation \cite{note2} -- for a series of
decreasing coupling strengths, (a)--(d), and assuming resonance with the
atom, Fig.~\ref{fig:four}, and a detuning from the atom, Fig.~\ref{fig:five}.
With the coupling strong compared to
the Einstein $A$ coefficient coherent Rabi oscillations are seen. There are
also discontinuous changes, which in the case of strong coupling are merely a
direct manifestation of the assumed quantum jumps for the atom. Note, however,
that the atom does not jump monotonously, ``up'' then ``down'' then ``up''
$\cdots$, as in Fig.~\ref{fig:two}; repeated up-jumps can
transfer many energy quanta to the field mode. At an intermediate coupling
strength, partial Rabi oscillations are still present. Once the coupling
becomes weak, though, the Rabi oscillations apparently disappear altogether,
and an entirely new kind of jump evolution sets in. These jumps proceed at
a rate far less than the overall jump rate for the atom. Their rate
decreases with the square of the coupling constant [(c) to (d)] and also
when the detuning is increased (from Fig.~\ref{fig:four} to
Fig.~\ref{fig:five}).

Having, then, assumed jumps for the atom in interaction with all but one of
the field modes, a jump evolution for the one remaining mode emerges naturally
in the weak-coupling limit. We close the self-consistent loop by showing that
the one mode samples a Bose-Einstein distribution, and by calculating the
rate of the single-mode jumps, to demonstrate that the sum over modes returns
the rates $\Gamma_{\rm up}$ and $\Gamma_{\rm down}$.

\subsection{Self-Consistency}
Let us denote the number of energy quanta shared between the atom and the
field mode at time $t$ by $n_\omega+1$. Let $t_k$ be the time of the very
last jump of the atom and $t_{k+1}$ be the time of the jump that is to occur
next. Then, for $t_k<t<t_{k+1}$, the entangled state of the atom and field
mode may be expanded as
\begin{equation}
|\bar\psi_{\rm REC}(t)\rangle=\bar C_{e|i}(t)|E_e\rangle|n_\omega\rangle
+\bar C_{g|i}(t)|E_g\rangle|n_\omega+1\rangle\;,
\end{equation}
with $\bar C_{e|i}(0)=\delta_{e,i}$ and $\bar C_{g|i}(0)=\delta_{g,i}$,
where $i$ is $e$ or $g$  for an up- or down-jump at $t_k$, respectively.
Writing
\begin{subeqnarray}
\bar C_{e|i}(t)&=&e^{-i(n_\omega+\frac12)\omega t}
e^{i\frac12\phi_{\hat{\vec n}}(\omega)}\tilde C_{e|i}(t)\;,\\
\bar C_{g|i}(t)&=&e^{-i(n_\omega+\frac12)\omega t}
e^{-i\frac12\phi_{\hat{\vec n}}(\omega)}\tilde C_{g|i}(t)\;,
\end{subeqnarray}
with $\phi_{\hat{\vec n}}(\omega)\equiv\arg[\kappa_{\lambda,\hat{\vec n}}
(\omega)]$, from (\ref{eqn:ten}) and (\ref{eqn:fifteen}), the equations of
motion for the conditional state amplitudes are
\begin{subeqnarray}
\frac{d{\tilde C}_{e|i}}{dt}&=&-{\textstyle{1\over2}\displaystyle}
(\Gamma_{\rm down}-i\Delta\omega)\tilde C_{e|i}-i|\kappa_{\lambda,\hat{\vec n}}(\omega)|
\sqrt{n_\omega+1}\,\tilde C_{g|i}\;,
\label{eqn:eighteena}\\
\frac{d{\tilde C}_{g|i}}{dt}&=&-{\textstyle{1\over2}\displaystyle}
(\Gamma_{\rm up}+i\Delta\omega)\tilde C_{g|i}-i|\kappa_{\lambda,\hat{\vec n}}(\omega)|
\sqrt{n_\omega+1}\,\tilde C_{e|i}\;.
\label{eqn:eighteenb}
\end{subeqnarray}

Consider now the case of an up-jump at $t_k$, such that the initial
state amplitudes are $\bar C_{e|i}(0)=1$ and $\bar C_{g|i}(0)=0$. For
weak coupling we will have $\bar C_{e|i}(t)\approx1$ and
$\bar C_{g|i}(t)\sim|\kappa_{\lambda,\hat{\vec n}}(\omega)|$, with an
overwhelming probability that the next jump of the atom will be a down-jump.
By a similar argument, it is highly likely that the down-jump is followed
by another up-jump; the most probable progression is then ``up'',
``down'', ``up'', $\ldots$, just as we illustrated it in Fig.~\ref{fig:two}
(notice that the number $n_\omega$ is unchanged throughout such a
progression). Due, however, to the small amplitude -- either $\bar C_{g|i}
(t)\sim|\kappa_{\lambda,\hat{\vec n}}(\omega)|$ or $\bar C_{e|i}(t)
\sim|\kappa_{\lambda,\hat{\vec n}}(\omega)|$ -- excited by the coupling
of the atom to the selected mode, there is always a small probability that
a jump will occur to break the alternating sequence. An up-jump might be
followed by a second up-jump or two down-jumps might occur in a row. These
events change $n_\omega$ and produce the jumps of the field mode seen in
Figs.~\ref{fig:four} and \ref{fig:five}. In Figs.~\ref{fig:four}(c) and
\ref{fig:five}(c), the presence of the small amplitude that underlies the
jump mechanism is still seen as a ``fuzz'' on top of the developing
smooth curve. The``fuzz'' is even present in Figs.~\ref{fig:four}(d)
and \ref{fig:five}(d), though there it is too small to be visible.

The physical interpretation of the anomalous events in the jump record of
the atom is that each represents the scattering of a photon between one
of the many field modes of the reservoir and the field mode selected to be
viewed. Two up-jumps occur in a row for example, because, in the interval
between them, the energy absorbed on the first jump is transferred to the
selected mode; the quantum trajectory resolves the transfer at the time of
the second up-jump.

Our task now is to calculate rates for the unlikely jumps. We do this using
the method of Sect.~IVD in \cite{carmichael97}. The equations of motion
(\ref{eqn:eighteena}) and (\ref{eqn:eighteenb}) give
\begin{subeqnarray}
\frac{d|\tilde C_{e|i}|^2}{dt}&=&-\Gamma_{\rm down}|\tilde C_{e|i}|^2
-2|\kappa_{\lambda,\hat{\vec n}}(\omega)|\sqrt{n_\omega+1}\,{\rm Im}
(\tilde C_{e|i}\tilde C_{g|i}^*)\;,
\label{eqn:nineteena}\\
\noalign{\vskip2pt}
\frac{d|\tilde C_{g|i}|^2}{dt}&=&-\Gamma_{\rm up}|\tilde C_{g|i}|^2
+2|\kappa_{\lambda,\hat{\vec n}}(\omega)|\sqrt{n_\omega+1}\,{\rm Im}
(\tilde C_{e|i}\tilde C_{g|i}^*)\;,
\label{eqn:nineteenb}\\
\noalign{\vskip2pt}
\frac{d{\rm Re}(\tilde C_{e|i}\tilde C_{g|i}^*)}{dt}&=&-{\textstyle{1\over2}
\displaystyle}(\Gamma_{\rm down}+\Gamma_{\rm up}){\rm Re}(\tilde C_{e|i}
\tilde C_{g|i}^*)-\Delta\omega{\rm Im}(\tilde C_{e|i}\tilde C_{g|i}^*)\;,
\label{eqn:nineteenc}\\
\noalign{\vskip2pt}
\frac{d{\rm Im}(\tilde C_{e|i}\tilde C_{g|i}^*)}{dt}&=&-{\textstyle{1\over2}
\displaystyle}(\Gamma_{\rm down}+\Gamma_{\rm up}){\rm Im}(\tilde C_{e|i}
\tilde C_{g|i}^*)+\Delta\omega{\rm Re}(\tilde C_{e|i}\tilde C_{g|i}^*)\cr
\noalign{\vskip2pt}
&&+|\kappa_{\lambda,\hat{\vec n}}(\omega)|\sqrt{n_\omega+1}\,
(|\tilde C_{e|i}|^2-|\tilde C_{g|i}|^2)\;.
\label{eqn:nineteend}
\end{subeqnarray}
We define
\begin{equation}
W_e\equiv\int_{t_k}^\infty\! dt\,|\tilde C_{e|i}(t)|^2\;,\qquad
W_g\equiv\int_{t_k}^\infty\! dt\,|\tilde C_{g|i}(t)|^2\;,
\end{equation}
and
\begin{equation}
U\equiv{\rm Re}\!\left[\int_{t_k}^\infty\!dt\,\tilde C_{e|i}(t)
\tilde C_{g|i}^*\right]\;,\qquad
V\equiv{\rm Im}\!\left[\int_{t_k}^\infty\!dt\,\tilde C_{e|i}(t)
\tilde C_{g|i}^*\right]\;,
\end{equation}
where $\Gamma_{\rm up}W_g$ and $\Gamma_{\rm down}W_e$ are the probabilities,
given $i$ is $e$ and $g$, respectively, that the unlikely jump will occur.
From the Laplace transforms of (\ref{eqn:nineteena}) -- (\ref{eqn:nineteend}),
we then have
\begin{subeqnarray}
-\delta_{e,i}&=&-\Gamma_{\rm down}W_e-2|\kappa_{\lambda,\hat{\vec n}}(\omega)
|\sqrt{n_\omega+1}\,V\;,
\label{eqn:twentytwoa}\\
\noalign{\vskip4pt}
-\delta_{g,i}&=&-\Gamma_{\rm up}W_g+2|\kappa_{\lambda,\hat{\vec n}}(\omega)
|\sqrt{n_\omega+1}\,V\;,
\label{eqn:twentytwob}\\
\noalign{\vskip4pt}
0&=&-{\textstyle{1\over2}\displaystyle}(\Gamma_{\rm down}+\Gamma_{\rm up})V
+\Delta\omega U+|\kappa_{\lambda,\hat{\vec n}}(\omega)|\sqrt{n_\omega+1}\,
(W_e-W_g)\;,
\label{eqn:twentytwoc}\\
\noalign{\vskip4pt}
0&=&-{\textstyle{1\over2}\displaystyle}(\Gamma_{\rm down}+\Gamma_{\rm up})U
-\Delta\omega V\;,
\label{eqn:twentytwod}
\end{subeqnarray}
and hence
\begin{subeqnarray}
\left.\Gamma_{\rm up}W_g\right|_{i=e}&=&\frac{\frac12(\Gamma_{\rm down}
+\Gamma_{\rm up})/\pi}{\left[\frac12(\Gamma_{\rm down}+\Gamma_{\rm up})\right]^2
+(\Delta\omega)^2}\frac{2\pi|\kappa_{\lambda,\hat{\vec n}}(\omega)|^2}
{\Gamma_{\rm down}}(n_\omega+1)\;,
\label{eqn:twentythreea}\\
\noalign{\vskip2pt}
\left.\Gamma_{\rm down}W_e\right|_{i=g}&=&\frac{\frac12(\Gamma_{\rm down}
+\Gamma_{\rm up})/\pi}{\left[\frac12(\Gamma_{\rm down}+\Gamma_{\rm up})\right]^2
+(\Delta\omega)^2}\frac{2\pi|\kappa_{\lambda,\hat{\vec n}}(\omega)|^2}
{\Gamma_{\rm up}}(n_\omega+1)\;,
\label{eqn:twentythreeb}
\end{subeqnarray}
where we have solved (\ref{eqn:twentytwoa})--(\ref{eqn:twentytwod}) to lowest
order in the coupling strength.

Equations (\ref{eqn:twentythreea}) and (\ref{eqn:twentythreeb}) specify the
probability for the unlikely jump to occur following any preparation of the
initial state $i$. To obtain photon-number jump rates, we must multiply by
the rate at which the state $i$ is prepared, i.e., by the jump rates for the
atom; (\ref{eqn:twentythreea}) is multiplied by $\Gamma_{\rm up}p_e^{\rm eq}$
 and (\ref{eqn:twentythreeb}) by $\Gamma_{\rm down}p_g^{\rm eq}$, where
$p_g^{\rm eq}$ and $p_e^{\rm eq}$ are the  state occupation probabilities in
thermal equilibrium.  We also set $n_\omega=N_\omega$ in (\ref{eqn:twentythreea})
and $n_\omega+1=N_\omega$ in (\ref{eqn:twentythreeb}), where $N_\omega$ is the
photon number expectation plotted in Figs.~\ref{fig:four} and \ref{fig:five}
(recall that $n_\omega+1$ is the number of quanta {\it shared\/} with the atom
at $t_k$). The photon-number jump rates are then
\begin{subeqnarray}
\gamma^{\rm up}_{N_\omega}&=&\frac{\frac12(\Gamma_{\rm down}
+\Gamma_{\rm up})/\pi}{\left[\frac12(\Gamma_{\rm down}+\Gamma_{\rm up})\right]^2
+(\Delta\omega)^2}2\pi|\kappa_{\lambda,\hat{\vec n}}(\omega)|^2(N_\omega+1)\,
p_e^{\rm eq}\;,
\label{eqn:twentyfoura}\\
\noalign{\vskip2pt}
\gamma^{\rm down}_{N_\omega}&=&\frac{\frac12(\Gamma_{\rm down}
+\Gamma_{\rm up})/\pi}{\left[\frac12(\Gamma_{\rm down}+\Gamma_{\rm up})\right]^2
+(\Delta\omega)^2}2\pi|\kappa_{\lambda,\hat{\vec n}}(\omega)|^2N_\omega\,
p_g^{\rm eq}\;.
\label{eqn:twentyfourb}
\end{subeqnarray}

The self-consistency of the thermal jump picture is now easy to demonstrate.
On the one hand, we use (\ref{eqn:twentyfoura}) and (\ref{eqn:twentyfourb})
to set up rate equations for the selected mode photon number and, using
detailed balance, solve these in steady state. Hence, we obtain the equilibrium
probability to find $N_\omega$ photons in the selected mode:
\begin{equation}
p_{N_\omega}^{\rm eq}=\left(1-p_e^{\rm eq}/p_g^{\rm eq}\right)\!\!
\left(\frac{p_e^{\rm eq}}{p_g^{\rm eq}}\right)^{\!\!N_\omega}\!
=[\bar n(\omega_0)+1]^{-1}\!\left[\frac{\bar n(\omega_0)}
{\bar n(\omega_0)+1}\right]^{N_\omega},
\end{equation}
where we have used $p_e^{\rm eq}/p_g^{\rm eq}=\Gamma_{\rm up}/\Gamma_{\rm down}$
and the Einstein formulas (\ref{eqn:sixa}) and (\ref{eqn:sixb}). We obtain
a Bose-Einstein distribution with average photon number $\bar N_\omega=
\bar n(\omega_0)$; note that the specific mode coupling strength and frequency
affects only the rate of approach to equilibrium. Of course, in reality, each
field mode couples to a vast number of two-state systems, and most strongly
to those with which it is nearly resonant. For the realistic situation we
would therefore find the expected $\bar N_\omega=\bar n(\omega)$, consistent
with the Planck radiation formula.

On the other side we must demonstrate the self-consistency of the jump rates.
To this end, we sum (\ref{eqn:twentyfoura}) and (\ref{eqn:twentyfourb}) over
all modes (all $\lambda$, $\hat{\vec n}$, $\omega$), with $N_\omega$ replaced
by its average value, and neglecting the frequency dependence of the density
of states and dipole coupling constant (in light of the Lorentzian resonance).
The resulting jump rates for the gain and loss of photons by the thermal
environment should equal the jump rates assumed initially for the atom. The
sums do, indeed, return $\Gamma_{\rm down}\,p_e^{\rm eq}$ and $\Gamma_{\rm up}
\,p_g^{\rm eq}$, showing that the net jump rates are in accord with the
Einstein rules (\ref{eqn:sixa}) and (\ref{eqn:sixb}) and Fermi's golden rule,
(\ref{eqn:seven}) and (\ref{eqn:eight}). This completes our demonstration
that thermal jump trajectories are consistent.

\section*{Acknowledgments}
This work was supported by the National Science Foundation under Grant
No.\ PHY-9531218 and by a Research Award of the Alexander von Humboldt-Stiftung.
HJC thanks Professor W. Schleich for his support and hospitality during his
stay at the University of Ulm.

%


\begin{thebibliography}{8.}
\bibitem{note1} It is interesting to note that Einstein motivated
his inclusion of stimulated emission jumps (as we would now call
them) along with absorption jumps by recalling just this
wave-based physics \cite{einstein16}: ``If a Planck resonator is
located in a radiation field, the energy of the resonator is changed
through the work done on the resonator by the electromagnetic field
of the radiation; this work can be positive or negative depending on
the phases of the resonator and the oscillating field. We
correspondingly introduce the following hypothesis. $\ldots$''
\bibitem{note2} This is to be understood in the sense of a true
expectation -- i.e., what is likely to be the case -- and not in the
sense of the quantum mechanical mean value.
\end{thebibliography}
\end{document}